\documentclass[a4paper]{article}

\usepackage{INTERSPEECH2022}

\usepackage{amsfonts,amsmath,amssymb,amsthm}
\usepackage{enumitem}
\usepackage{graphicx}
\usepackage{hyperref}
\usepackage[utf8]{inputenc}
\usepackage{mathtools}
\usepackage{microtype}
\usepackage{moresize}
\usepackage{multirow}
\usepackage{siunitx}
\usepackage{xcolor}
\usepackage{xparse}
\usepackage{xspace}

\definecolor{mygreen}{HTML}{11ab11}
\definecolor{myblue}{HTML}{7373df}
\newcounter{NumberOfTODOs}

\newcommand{\augmentationfunc}{f}
\newcommand{\augmentationparam}{{\pmb \theta}}
\newcommand{\augment}[1]{\augmentationfunc(#1, \augmentationparam)}
\newcommand{\augmentA}[1]{\augmentationfunc(#1, \augmentationparam_A)}
\newcommand{\augmentB}[1]{\augmentationfunc(#1, \augmentationparam_B)}
\newcommand{\bitrate}{R}
\newcommand{\cleanwaveform}{\waveform^{c}}
\newcommand{\codebooksize}{N}

\newcommand{\confintsym}[2]{#1 \scriptstyle{\, \pm \, #2}}
\newcommand{\dec}{{\rm dec}}

\newcommand{\embedding}{\textbf{z}}
\newcommand{\embeddingpartone}{\embedding^{(1)}}
\newcommand{\embeddingparttwo}{\embedding^{(2)}}
\newcommand{\enc}{{\rm enc}}

\newcommand{\frameindex}{t}
\newcommand{\framerate}{R_f}

\newcommand{\LOne}{L_1}
\newcommand{\LTwo}{L_2}
\newcommand{\loss}{\mathcal{L}}

\newcommand{\lossgen}{\loss}
\newcommand{\lossgenrec}{\lossgen_{\rm rec}}

\newcommand{\norm}[1]{\left\|#1\right\|}

\newcommand{\numfeatures}{D}
\newcommand{\numframes}{F}
\newcommand{\numquantizerlayers}{n_q}
\newcommand{\numsamples}{T}
\newcommand{\reals}{\mathbb{R}}
\newcommand{\reconwaveform}{\hat{\waveform}}
\newcommand{\spec}{\mathcal{S}}
\newcommand{\specframe}{\spec^\windowlength_\frameindex}
\newcommand{\subfigcaption}[1]{\textbf{#1)}}
\newcommand{\swappedwaveform}{\widetilde{\waveform}}
\newcommand{\SoundStream}{\emph{SoundStream}\xspace}
\newcommand{\tsixty}{T_{60}}

\newcommand{\visqol}{ViSQOL\xspace}
\newcommand{\waveform}{\textbf{x}}
\newcommand{\windowlength}{s}

\title{Disentangling speech from surroundings with neural embeddings}
\name{Ahmed Omran, Neil Zeghidour, Zal\'an Borsos,\\ F\'elix de~Chaumont~Quitry, Malcolm Slaney, Marco Tagliasacchi}
\address{Google Research}
\email{\{ahmedomran,neilz,zborsos,fcq,malcolmslaney,mtagliasacchi\}@google.com}

\begin{document}

\maketitle

\begin{abstract}
    We present a method to separate speech signals from noisy environments in the embedding space of a neural audio codec. We introduce a new training procedure that allows our model to produce structured encodings of audio waveforms given by embedding vectors, where one part of the embedding vector represents the speech signal, and the rest represent the environment. We achieve this by partitioning the embeddings of different input waveforms and training the model to faithfully reconstruct audio from mixed partitions, thereby ensuring each partition encodes a separate audio attribute. As use cases, we demonstrate the separation of speech from background noise or from reverberation characteristics. Our method also allows for targeted adjustments of the audio output characteristics.
\end{abstract}

\section{Introduction}

Recent progress in generative models of audio have allowed for the development of neural codecs \cite{morishima1990,kankanahalli2018,garbacea2019,zeghidour2021a} that are competitive with traditional codecs. However, while traditional codecs such as Opus~\cite{valin2012} cascade blocks with specific roles (e.g., voice activity detection, noise shaping), neural codecs typically stack learned representations that are not easily interpretable. This can be detrimental to downstream tasks, such as audio event detection, and selective transmission of parts of the signal --- such tasks can potentially benefit from disentangled representations in the compressed domain, where different attributes of the signal are represented separately.

In this work, we train a neural audio codec based on \SoundStream~\cite{zeghidour2021a} to represent audio signals in a structured way in the compressed domain. The neural network maps input audio into an embedding space, which we split into partitions that are each to capture a different attribute of the input audio. We then use targeted augmentations of the data as well as a custom loss function to introduce a strong inductive bias in the model to specifically allocate each partition to a given attribute. The compressed representations can then be tailored to make best use of the available bit rate, by prioritizing certain types of content over others, or only transmitting the desired part of the signal, such as the clean speech content.

We demonstrate two different types of disentangling tasks: The first is separating speech from background noise. The noise signal is an additive component that varies on similar time scales as speech changes. But in contrast to conventional source separation and speech enhancement~\cite{loizou2007,kavalerov2019,wisdom2020}, the separation happens in the embedding space. The second is disentangling speech from the reverberation characteristics of the recording environment. Here, the reverberation signal cannot be simply subtracted from the original signal in the time domain, and is instead modeled as a convolution with a time-invariant room impulse response that spans an extended period of time. An important property we exploit in our method is that the reverberation characteristics typically change more slowly than the speech signal. For both cases, we demonstrate the separation of speech from the environment in the embedding space, and how this disentanglement enables fine-grained control over the synthesized output audio.

\begin{figure*}
    \centering
    \includegraphics[width=\textwidth]{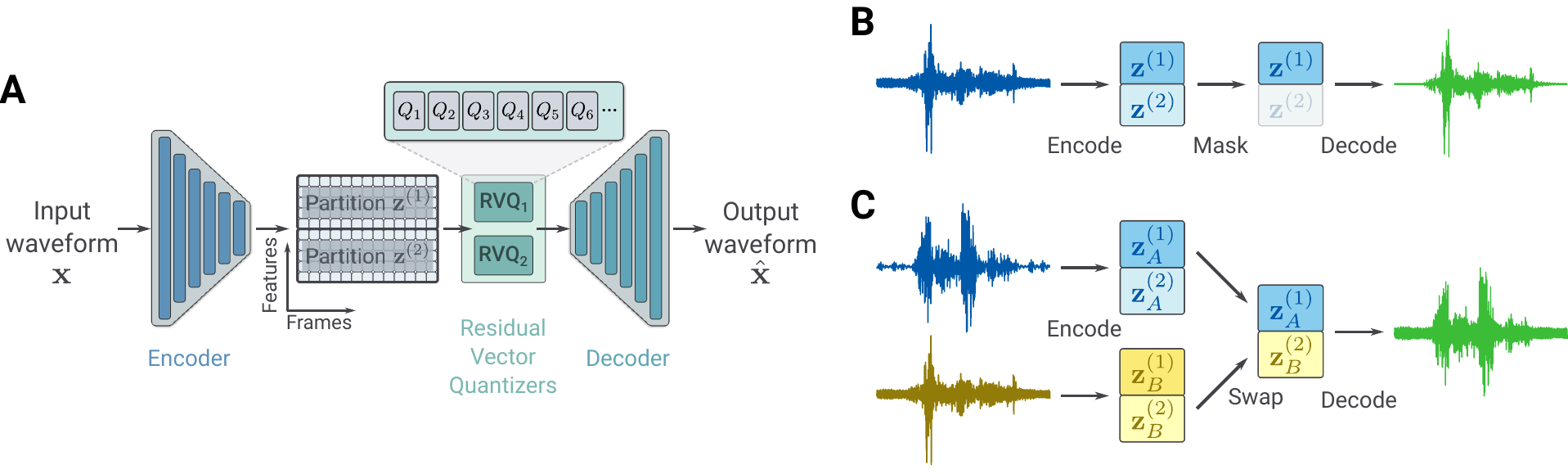}
    \caption{The architecture of our model. \subfigcaption{A} An input waveform is processed by the encoder to produce one embedding vector per frame. Each partition is processed using a separate residual vector quantizer. The quantized embeddings are converted back into an audio waveform. \subfigcaption{B} The embedding partitions can be individually set to zero before decoding, which removes the corresponding attribute from the output audio, e.g., the background noise. \subfigcaption{C} An audio waveform can be encoded and an embedding partition swapped out with that of a different waveform, thereby replacing an audio attribute of the first waveform with that of the second.}
    \label{fig:architecture}
\end{figure*}

\section{Related work}

A standard approach for disentangling attributes relies on learning a factorized latent space of a generative model~\cite{higgins2018}, e.g., in the bottleneck of an autoencoder. Several works have used an autoencoder with a classifier that tries to predict an attribute from the embeddings, while the encoder is trained to fool the classifier by generating embeddings that are invariant under a change in this attribute. This has been applied to disentanglement of image features~\cite{lample2017}, sensor anonymization~\cite{hajihassnai2021}, voice conversion~\cite{chou2018,nguyen2021} and music translation~\cite{mor2018}. This idea has also been proposed for disentangling multiple speech attributes by chaining several autoencoders trained in this fashion~\cite{gong2018}. However, this adversarial training adds complexity and potential instability, and the necessary labels are difficult to obtain in the audio domain. Here, we instead rely on a suitable choice of input and target data and training objectives to get sufficient inductive biases, which are necessary for disentanglement~\cite{locatello2020}.

To avoid issues of adversarial training, some works rely on constraining the information capacity by a careful choice of embedding dimensions that ensures each channel contains only one factor of variation, such as AutoVC~\cite{qian2019} and SpeechSplit~\cite{qian2020a}. However this comes with a trade-off in the generated output quality, and does not provide flexibility in choosing the embedding dimension for a signal attribute of interest, whereas our method provides more freedom in choosing the embedding dimensions.

Polyak et al.~\cite{polyak2021} used self-supervised learning to train multiple encoders to obtain discrete representations of speech inputs, each one capturing a different speech attribute. These representations are resynthesized to high-quality speech, enabling very low-bitrate speech transmission. This is only limited to speech characteristics and relies on domain-specific encoders, whereas our codec can process a broader audio domain. Yang et al.~\cite{yang2021} demonstrated source separation and neural coding in the latent space, an approach that is only limited to signals that can be added together in the time domain.

Disentangled representations can be learned by partitioning feature vectors in the latent space and mixing the partitions while training the model to produce realistic decoded outputs, thereby capturing one factor of variation in each partition~\cite{hu2018} or to synthesize new samples that share attributes between different input samples~\cite{park2020}. A related approach is to supply input data that differs in some attribute and constrain different subspaces of the latent vectors to capture those changes, as was shown in the vision domain with dataset batching~\cite{kulkarni2015} or targeted augmentations~\cite{bhagat2020}, or for the SPICE audio pitch estimator~\cite{gfeller2020}. Inspired by these mixing approaches, we present how to separate speech signals from their environment, and demonstrate disentanglement of speech from background noise or reverberation in a way that enables fine-grained control over the output.

\section{Methods}

Our model is based on \SoundStream\cite{zeghidour2021a}, a streamable neural audio codec with a vector-quantized autoencoder architecture. However, unlike the baseline \SoundStream model, we have the following additional requirements: We need to set the dimension allocated to each embedding partition, and for compression we need a way of setting the effective bitrate for each channel. The codec should allow for streaming audio while producing structured embeddings in real time, such that certain types of content can have their bitrate reduced or be removed entirely on demand.

\subsection{Model}

Our model architecture is shown in Fig.~\ref{fig:architecture}A. Single-channel audio waveforms $\waveform \in \reals^\numsamples$ are processed by the encoder, which is a convolutional network with blocks consisting of residual units and a strided convolutional layer. All convolutions have causal padding to allow for real-time streaming of the input waveforms. The encoder produces embeddings $\embedding = \enc(\waveform) \in \reals^{\numframes\times\numfeatures}$, where $\numframes$ is the number of frames, and $\numfeatures$ is the number of features per frame, set by the number of convolutional filters at the bottleneck. The number of frames $\numframes$ is given by the number of waveform samples $\numsamples$ divided by the product of all convolution striding factors.

We apply residual vector quantization to each embedding partition separately. A residual vector quantizer~\cite{zeghidour2021a} is a stack of vector quantization layers, each of which replaces its input by a vector from a learned discrete codebook and passes the residual error to the next layer. The effective bitrate is given by $\bitrate = \framerate \codebooksize \numquantizerlayers$, where $\framerate$ is the frame rate, $\codebooksize$ is the codebook size in bits, and $\numquantizerlayers$ is the number of quantizer layers. Each embedding partition, a subset of the $\numfeatures$ total dimensions, is processed by such a quantization module, allowing us to set a separate bitrate for each partition, particularly when different attributes change over different time scales. For a finite bitrate budget, we can also prioritize speech content over other attributes like the room characteristics.

The quantized embeddings are concatenated together along the feature axis and fed into the decoder, which converts them into a reconstructed audio waveform $\reconwaveform = \dec(\embedding) \in \reals^\numsamples$. The decoder is very similar in structure to the encoder, where convolutions are replaced by transposed convolutions~\cite{zeghidour2021a}.

\subsection{Disentanglement scheme}
\label{subsec:scheme}

\begin{table*}[t]
    \centering
    \caption{Audio quality after decoding the full embeddings and individual partitions as measured with \visqol, where the expected target audio is used as a reference, that is, noisy speech, clean speech and noise, respectively. Values are averaged over $250$ noisy waveforms of $\SI{3}{\second}$ duration. Uncertainties denote $95\%$~confidence intervals.}
    \renewcommand{\arraystretch}{1.2}
    \begin{tabular}{c|c c c}
        \hline\hline
        \multirow{3}{*}{\visqol} & \multirow{3}{*}{\shortstack{Multichannel\\quantization\\($\SI{6.3}{\kilo bps}$/channel)}} & \multirow{3}{*}{\shortstack{Single channel\\quantization\\($\SI{12}{\kilo bps}$)}} & \multirow{3}{*}{No quantization}\\ \\ \\
        \hline\hline
        Reconstruction  & $\confintsym{3.88}{0.03}$  & $\confintsym{3.84}{0.04}$  & $\confintsym{4.32}{0.03}$ \\ 
        Speech partition& $\confintsym{3.16}{0.08}$  & $\confintsym{3.08}{0.08}$  & $\confintsym{3.06}{0.08}$ \\
        Noise partition & $\confintsym{2.54}{0.05}$  & $\confintsym{2.55}{0.06}$  & $\confintsym{2.57}{0.05}$ \\
        \hline\hline
    \end{tabular}
    \label{tab:noise_disentanglement_visqol}
\end{table*}

Disentangling speech from its environment requires careful design of the training objectives, in order to both achieve high-quality audio transmission and create structured embeddings with no content leakage between different partitions.

We apply an augmentation $\augmentationfunc(\cdot, \augmentationparam)$ to a clean speech waveform $\cleanwaveform$ to produce $\waveform = \augment{\cleanwaveform}$. Here we consider parameters $\augmentationparam$ that are used as additive signals to the clean speech $\waveform = \cleanwaveform + \augmentationparam$, or as convolution kernels $\waveform(t) = \int \augmentationparam(\tau)\cleanwaveform(t-\tau) {\rm d}\tau$. The aim is to produce embeddings where $\cleanwaveform$ is encoded in a dedicated partition $\embeddingpartone$, and $\augmentationparam$ is encoded in a separate partition $\embeddingparttwo$, such that $\embedding = {\rm concat}(\embeddingpartone, \embeddingparttwo)$. Additionally, we require a good reconstruction quality of the audio and speech signal of interest. Therefore, during each training iteration we create three different types of augmented signals, which are used sequentially as input to the training objective in Section~\ref{subsec:objectives}:

\begin{enumerate}[label=(\roman*)]
\item Reconstruct the augmented input audio $\waveform$ using the full embedding vectors $\embedding$.
\item Reconstruct the clean speech component $\cleanwaveform$ from the augmented input $\waveform$. Here, the embeddings are multiplied by a binary mask to set the second partition $\embeddingparttwo$ that encodes the augmentation to zero (Fig.~\ref{fig:architecture}B).
\item Encode two different augmented input waveforms $\waveform_A = \augmentA{\cleanwaveform_A}$ and $\waveform_B = \augmentB{\cleanwaveform_B}$, where $\augmentationparam_A$ and $\augmentationparam_B$ represent different augmentations. This produces embeddings $\embedding_A$ and $\embedding_B$, which are partitioned as $\big[\embeddingpartone_A, \embeddingparttwo_A\big]$ and $\big[\embeddingpartone_B, \embeddingparttwo_B\big]$, respectively. Then, the second partition of the embedding vectors is swapped, and the new embedding vectors $\big[\embeddingpartone_A, \embeddingparttwo_B\big]$ are decoded to reconstruct a new version $\swappedwaveform_A = \augmentB{\cleanwaveform_A}$ that inherits the attribute of interest from $\waveform_B$ (Fig.~\ref{fig:architecture}C).
\end{enumerate}

\subsection{Training objectives}
\label{subsec:objectives}

To train the model, we minimize a multi-scale spectral reconstruction loss for each step~\cite{engel2020,gritsenko2020}:
\begin{equation}
\begin{aligned}
    \lossgenrec(\waveform, \waveform') = &\sum_{\log_2 \windowlength = 6}^{11}\bigg( \sum_\frameindex \norm{\specframe(\waveform) - \specframe(\waveform')}_1 + \\ &\sqrt{\frac{\windowlength}{2}} \sum_\frameindex \norm{\log \specframe(\waveform) - \log \specframe(\waveform')) }_2 \bigg),
\end{aligned}
\label{eq:ddsp_loss}
\end{equation}
where $\specframe(\cdot)$ denotes the $\frameindex$-th frame of a 64-bin mel-spectrogram with a window length of $\windowlength$ and hop length of $\windowlength/4$. The $\LOne$ objective in the first term encourages matching the highest-energy features of the signal, while the $\LTwo$ logarithmic term provides attention to the quieter details. For each of the three training steps, we compute this loss function between the reconstructed waveform $\reconwaveform$ and the corresponding target waveform, i.e. $\lossgenrec(\waveform, \reconwaveform)$ for training step (i), $\lossgenrec(\cleanwaveform, \reconwaveform)$ for step (ii), and $\lossgenrec(\swappedwaveform, \reconwaveform)$ for step (iii).

When relying only on the reconstruction losses~\eqref{eq:ddsp_loss}, we find that the output audio suffers from robotic artifacts. To alleviate these, we use a set of discriminators operating on the audio waveforms at multiple scales~\cite{zeghidour2021a}, and train them in an adversarial fashion to encourage the \SoundStream model to produce more realistically sounding audio. In contrast to previous work, we do not use discriminators operating in the STFT domain, as we found they add no performance benefit, and instead our discriminators operate on multiple scales on the time-domain audio. In addition to minimizing the adversarial loss, we also rely on a feature loss from the absolute differences of the discriminator layer outputs between the original and reconstructed audio~\cite{zeghidour2021a}. However, we observed that simultaneous training of all components often yields poor output quality and find better results by first training the model on all three tasks using reconstruction losses only, then freezing the parameters of the encoder and quantizer, and finally training the decoder alongside the discriminators to improve the output quality. The optimizers, learning rate and loss weights are identical to those used for tranining \SoundStream~\cite{zeghidour2021a}.

\section{Results}

\begin{figure*}
    \centering
    \includegraphics[width=0.95\textwidth]{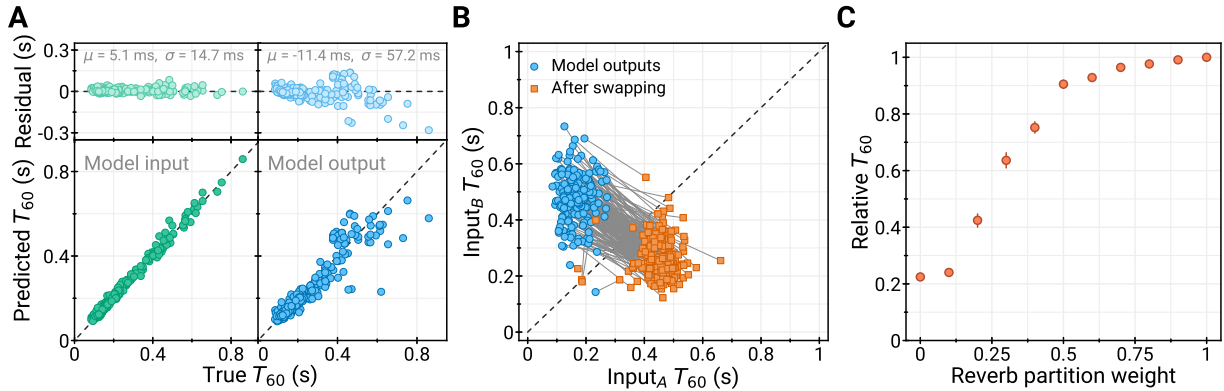}
    \caption{\subfigcaption{A} Benchmark of the $\tsixty$ estimator and our model's reconstruction fidelity. In the lower plots, the classified values are compared with the true $\tsixty$ values known from the augmentation, both for the model inputs and reconstructed outputs, where data points with a perfect reproduction of the $\tsixty$ value should lie on the diagonal dashed line. The upper plots show the residuals, the means of the error and their standard deviations. \subfigcaption{B} $\tsixty$ values of pairs of inputs after reconstruction by the model (blue circles) and with a swapped reverb embedding partition (orange squares). Solid gray lines connect points that belong to the same pair of inputs. \subfigcaption{C} Demonstration of reverb tuning. $200$ strongly reverberant waveforms ($\tsixty > \SI{400}{\milli\second}$) are encoded, and the reverb partition is multiplied by a global weight factor before decoding. The measured $\tsixty$ values are normalized by the value at weight factor of 1 and then averaged. Error bars denote $95\%$ confidence intervals.
    }
    \label{fig:reverb_disentanglement}
\end{figure*}

\subsection{Separating speech from noise}

We use speech waveforms from the LibriVox dataset~\cite{kearns2014}, and synthesize noisy speech waveforms by mixing these with noise from the Freesound dataset~\cite{fonseca2017}, which was screened to remove speech content and only keep clips with a CC0 license. Both the speech and noise waveforms are normalized, then the noise is multiplied by a random gain drawn from a normal distribution with mean of $\SI{-5}{\decibel}$ and standard deviation of $\SI{10}{dB}$ before being mixed with the speech waveform. In the input pipeline, we supply pairs of waveforms $(\waveform_A, \cleanwaveform_A)$ and $(\waveform_B, \cleanwaveform_B)$, each with the noisy and clean version, and where $\waveform_A$ and $\waveform_B$ differ in speech and background noise signals.

We encode each audio frame of $320$ samples into a $256$-dimensional embedding vector, which is partitioned into two equal halves, one to carry information about the speech and the other about background noise. Each embedding channel is processed with a dedicated residual vector quantizer, with $\numquantizerlayers=14$ layers of $\log_2 \codebooksize=9$ bits of depth each. Embedding vectors are produced with a frame rate of $\framerate=\SI{50}{\hertz}$, leading to an effective bitrate per channel of $\bitrate=\SI{6.3}{\kilo bps}$. We then train the model on the three objectives detailed in Section~\ref{subsec:scheme} using waveforms $\waveform_A$ and $\waveform_B$.

Table~\ref{tab:noise_disentanglement_visqol} shows the reconstruction quality of the model as measured with \visqol~\cite{hines2015,chinen2020}, an objective quality metric that correlates well with subjective evaluations. We show the scores for the reconstruction of the full embeddings, each individual partition, and the reconstructed waveform from mixed embeddings of two separate inputs. There, our trained model with separate quantization of each channel is compared with two other models: One with a global residual vector quantizer operating on all embeddings with a bitrate of $\bitrate=\SI{12}{\kilo bps}$, and one where no vector quantization is performed. The model with individually quantized channels performs better on average than the one with a global quantization of the embeddings. The model without quantization yields better quality when reconstructing the noisy speech signal owing to the absence of quantization noise, but it offers no signal compression, and the clean speech reconstruction is similar to the other models.

When reconstructing clean speech from the first embedding partition, the background noise is efficiently removed without substantially hurting speech intelligibility,\footnotemark but broadband stationary noise in the waveform is not fully suppressed, limiting the values of quantitative metrics like \visqol. We observe that decoding the noise embedding partition while setting the speech embeddings to zero leads to a reconstruction of the noise component, albeit at lower quality as this reconstruction was not an explicit training task. We also find that multiplying a partition with a weight factor between $0$ and $1$ leads to a reduction in volume of the corresponding audio content, a feature that can be used for adjusting the target level of denoising.\footnotemark[\value{footnote}]

\footnotetext{Audio examples at \url{https://google-research.github.io/seanet/disentangling_soundstream/examples/}}

\subsection{Separating speech from reverberation}

Synthetic reverberated speech differs from speech with background noise in that reverberation is not an additive signal, but rather the room impulse response is convolved with the speech. Another difference is that the information about the room impulse response typically changes much more slowly than the underlying speech content, and for the purpose of this work is assumed to be time-invariant.

We can use the same general framework to train a new codec that separates speech from reverberation. Here we allocate 54 embedding features to speech, which proved sufficient for this purpose. To capture the difference in temporal variations speech and reverberation, we modify the encoder to include an additional convolutional layer striding over multiple frames that outputs a second set of $10$-dimensional embedding vectors at a $10$-times lower frame rate than the speech embeddings, to capture the typical time scales of the room impulse response. We quantize the reverberation partition with only $\numquantizerlayers=4$ layers of $\log_2 \codebooksize=9$ bit depth, yielding an effective bitrate of $\bitrate=\SI{180}{bps}$. These ``slow'' embeddings allocated for reverberation are then upsampled by a factor of $10$ and concatenated with the ``fast'' speech embeddings along the feature axis.

We train on clean speech waveforms from the LibriVox dataset that are each augmented with a time-invariant, synthetic room impulse response, which represents a stationary speaker and receiver. The room impulse responses have characteristic $\tsixty$ times between $\SI{70}{\milli\second}$ and $\SI{1.2}{\second}$ with an average of $\SI{214}{\milli\second}$ and median of $\SI{272}{\milli\second}$. In the input pipeline, we provide pairs of waveforms ($\waveform_A$, $\cleanwaveform_A$) and $(\waveform_B, \swappedwaveform_B)$, where $\cleanwaveform_A$ is the non-reverberant version of $\waveform_A$, and $\swappedwaveform_B = \augmentA{\cleanwaveform_B}$ is the version of $\waveform_B$ that has the same room impulse response as $\waveform_A$. To make the task more explicit, $\waveform_A$ is augmented with strong reverberation levels ($\tsixty(\waveform_A)\geq \SI{400}{\milli\second}$) and $\waveform_B$ with weaker reverberation ($\tsixty(\waveform_B) \leq \SI{250}{\milli\second}$). We then train on the three objectives detailed in Section~\ref{subsec:scheme}.

We measure the reverberation characteristics of the waveforms using a pretrained estimator of $\tsixty$, the time reverberated sound takes to decay by \SI{60}{\decibel}. Its accuracy is shown in Fig.~\ref{fig:reverb_disentanglement}A. Input waveforms $\waveform$ are fed to the estimator, which yields estimated $\tsixty$ values that do not differ significantly from the ground-truth values used in the augmentation. We also evaluate the reconstructed waveforms $\reconwaveform$ generated by our model, and find a faithful reconstruction of the reverberation characteristics except for very long reverberation times.

To verify the separation of reverberation from speech in the embedding space, we input pairs of audio waveforms $(\waveform_A, \waveform_B)$ to our model to obtain reconstructed waveforms $(\reconwaveform_A, \reconwaveform_B)$ and evaluate their $\tsixty$ values. We then take the encoded embeddings of both waveforms and swap their reverberation embedding partitions before decoding and estimate the $\tsixty$ values of the new waveforms ($\swappedwaveform_A, \swappedwaveform_B)$. Fig.~\ref{fig:reverb_disentanglement}B shows the $\tsixty$ values before and after this swapping operation, where most pairs have their $\tsixty$ values successfully swapped, indicating the second partition includes all the information about the reverberation.

We further demonstrate the tunability of the decoded outputs by encoding $200$ strongly reverberated examples ($\tsixty \geq \SI{400}{\milli\second}$), multiplying the partition encoding reverberation by a weight factor between $0$ and $1$ and estimating the decoded output's $\tsixty$ time (Fig.~\ref{fig:reverb_disentanglement}C). We observe that the weight factor smoothly reduces $\tsixty$ in a predictable way, indicating that we can fine-tune the reverberation of speech on demand.

\section{Conclusion}

In this work, we introduce a training scheme to separate speech from background noise or the room reverberation in the embedding space of a neural audio codec. By alternating between reconstructing input audio, reconstructing only the speech component, and reconstructing new synthetic audio by decoding different combinations of embedding partitions, we achieve effective disentanglement of these contents. We further show how the separation in embedding space leads to a tunable output, for example, by removing the noise component altogether, tuning out the room reverberation, or inheriting a different room impulse response from a separate reference waveform. In future studies, we will scale up this scheme to disentangle multiple attributes at the same time and extend our approach to other factors of variation like pitch and speaker identity.

\section{Acknowledgments}

The authors thank Olivier Bachem, Hannah Muckenhirn, John Hershey, Ben Laurie, Dominik Roblek, Beat Gfeller, Yunpeng Li for helpful discussions, and Dan Ellis and Dick Lyon for technical discussions and helpful feedback on the manuscript.

\bibliographystyle{IEEEtran}
\bibliography{main}

\end{document}